\documentclass[prd,twocolumn,showpacs,preprintnumbers,aps,nofootinbib]{revtex4-1}
\usepackage{graphicx}
\usepackage{dcolumn}
\usepackage{bm}
\usepackage{amsmath,amssymb,amsfonts}
\usepackage{latexsym}
\usepackage{color}
\def\dcp{\Delta \mathcal{A}_{\text{CP}}}
\def\bsg{\mathcal{B}(B\to X_s \gamma)}
\def\rbb{\rho_{bb}}
\def\rtt{\rho_{tt}}
\def\irbb{\mbox{Im}(\rho_{bb})}
\def\rrbb{\mbox{Re}(\rho_{bb})}
\def\fbi{fb$^{-1}$}
\def\cba{c_{\beta-\alpha}}

\begin{document}

\preprint{CTPU-PTC-18-37}

\title{Electroweak baryogenesis via bottom transport}

 \author{Tanmoy Modak$^1$}
 \email{tanmoyy@hep1.phys.ntu.edu.tw}
 \author{Eibun Senaha$^2$} 
 \email{senaha@ibs.re.kr}
 \affiliation{$^1$Department of Physics, National Taiwan University, Taipei 10617, Taiwan}
\affiliation{$^2$Center for Theoretical Physics of the Universe, Institute for Basic Science (IBS), Daejeon 34126, Korea}
\bigskip


\begin{abstract}
We consider a scenario in which an extra bottom Yukawa coupling can drive electroweak baryogenesis
in the general two-Higgs doublet model. 
It is found that the new bottom Yukawa coupling with $\mathcal{O}(0.1)$ in magnitude can generate
the sufficient baryon asymmetry without conflicting existing data.
We point out that future measurements of the bottom Yukawa coupling at High-Luminosity
Large Hadron Collider and International Linear Collider, together with the CP asymmetry 
of $B\to X_s\gamma$ at SuperKEKB provide exquisite probes for this scenario.
\end{abstract}

\maketitle

\section{Introduction}
Existence of the baryon asymmetry of the Universe (BAU) is firmly established by various 
cosmological observations such as the cosmic microwave background and big-bang nucleosynthesis~\cite{Tanabashi:2018oca}.  
However, its origin is still unclear, which motivates one to search for physics beyond the Standard Model (SM).

A plethora of baryogenesis scenarios have been proposed so far.
After the discovery of the Higgs boson at Large Hadron Collider (LHC) in 2012~\cite{125h}, a significant 
attention has been paid in particular to electroweak baryogenesis (EWBG)~\cite{ewbg,Morrissey:2012db} for its close
connection to Higgs physics. One of the necessary conditions for the successful EWBG is that electroweak
phase transition (EWPT) is strongly first order, which requires an extra particle
with a mass of sub-TeV that couples to the Higgs boson. 
Well-studied examples are new scalar or vector particles that modify the Higgs potential 
by tree-level mixings and/or loop effects.
In addition to this conventional cases, it is pointed out that even fermions 
could induce such effect if they couple to the Higgs boson strongly~\cite{Carena:2004ha} 
(for a recent study, see Ref.~\cite{Huang:2016cjm}). But in this case, additional bosonic degrees of freedom
are needed to evade vacuum instability.

Furthermore, CP violation relevant to EWBG often arises from Higgs-Yukawa interactions.
Therefore, the Higgs signal strengths are inevitably modified by the new physics (NP) effects.

Recently, the Higgs boson decay to bottom quarks has been observed at the LHC.
Its signal strength relative to the SM expectation is 
$1.01\pm 0.12(\text{stat.})^{+0.16}_{-0.15}(\text{syst.})$ at ATLAS~\cite{Aaboud:2018zhk}
and $\mu=1.04\pm 0.14(\text{stat.})\pm 0.14(\text{syst.})$ at CMS~\cite{Sirunyan:2018kst}, respectively.
While the measured values are consistent with the SM, however, there still exist sufficient room for NP.

The NP effects in the bottom sector is of great importance for $B$ physics as well.
In addition to the on-going LHCb experiment, Belle-II at KEK will start collecting data (phase 3) 
in early 2019 and accumulate it up to 50 $\text{ab}^{-1}$ by 2024.
One of the goals is to search for CP violation beyond the Cabibbo-Kobayashi-Maskawa (CKM) framework~\cite{CKM}.
It is of broad interest whether such a CP violation can be related to the BAU.

In this paper, we consider a scenario in which additional bottom Yukawa coupling is
responsible for the BAU and discuss its implications to collider phenomenology as well as $B$ physics,
especially $B\to X_s\gamma$. We take the general two Higgs double model (G2HDM)~\cite{Branco:2011iw} as a benchmark model.
For previous studies of EWBG in the model, see, {\it e.g.}, Refs.~\cite{Tulin:2011wi,Cline:2011mm,Guo:2016ixx,Liu:2011jh,Chiang:2016vgf,Fuyuto:2017ewj,deVries:2017ncy}. 
For instance, in Ref.~\cite{Fuyuto:2017ewj} a scenario in which BAU is sourced by
new CP violation in the up-type Yukawa couplings is considered. This EWBG scenario
is very efficient as long as an extra top Yukawa coupling is complex and $\mathcal{O}(0.1\text{-}1)$ in magnitude.
In such a case, there is no strong motivation to consider additional CP violation in the down-type Yukawa couplings.
In the current analysis, however, we explore the EWBG possibility assuming that the 
up-type Yukawa couplings do not provide any new CP violation. Therefore, the current 
analysis is complementary to the above top-driven scenario. While it is also discussed
that the CP violation required to generate BAU may also come from a flavor-changing bottom Yukawa coupling~\cite{Liu:2011jh} and evade 
the constraint from electric dipole moments (EDMs) of electron,
however, it is well known fact that the flavor-diagonal Yukawa couplings
are much more efficient~\cite{Cline:2011mm,Guo:2016ixx,Chiang:2016vgf,Fuyuto:2017ewj}.
Therefore, such possibilities should also be clarified in the bottom transport scenario. 
In contrast to the claim of Ref.~\cite{Liu:2011jh}, we cannot find any successful EWBG regions
utilizing the flavor-changing bottom Yukawa coupling in our numerical analysis.

We point out that the extra flavor-diagonal bottom Yukawa coupling of $\mathcal{O}(0.1)$ in magnitude 
can offer the successful EWBG without upsetting existing experimental constraints. 
It is found that, except some corner of the parameter space, most EWBG-viable regions can fully be 
covered by Higgs signal strength measurements at High-Luminosity LHC (HL-LHC) and future colliders such as International Linear
Collider (ILC). Besides, such scenario can also be tested by $B$ physics observables, especially the branching ratio and 
CP asymmetry of $B\to X_s\gamma$ at SuperKEKB. 

The paper is organized as follows. In Sec.~\ref{bau} we discuss about the formalism for EWBG via bottom transport.
Sec.~\ref{expc} is dedicated for the experimental constraints on the parameter space which is relevant for baryogenesis.
The results are summarized in Sec.~\ref{res}, with some discussions and conclusion in Sec.\ref{discu}.

\section{BAU via bottom transport}\label{bau}
The Yukawa interactions of the G2HDM in a generic basis are parametrized as 
\begin{align}
-\mathcal{L}_Y
\color{red}&=\bar{f}_L(Y_1\Phi_1+Y_2\Phi_2)f_R +\text{H.c},
\end{align}
where $\Phi_{1,2}$ are the Higgs doublets whose vacuum expectation values (VEVs)
are parametrised as $v_1$ and $v_2$, respectively and $f=u,d,e$.
In the mass eigen-basis of the fermions and Higgs bosons, one has
\begin{align}
-\mathcal{L}_Y&\ni \bar{f}_{L}y_{\phi}^ff_{R}\phi+\bar{u}\Big[V\rho^dP_R-\rho^{u\dagger}VP_L\Big]dH^++\text{H.c},
\end{align}
where $P_{L,R}=(1\mp \gamma_5)/2$, $V$ is the CKM 
matrix, $H^+$ is charged scalar and $\phi=h,H,A$, with $h$ is identified as 125 GeV boson, 
and $H$ and $A$ are CP-even and CP-odd scalars respectively.
$y_\phi^f$ are the $3\times 3$ matrices defined, respectively, as
\begin{align}
y_{hij}^f&= \frac{\lambda_i^f}{\sqrt{2}}\delta_{ij}s_{\beta-\alpha}+\frac{\rho_{ij}^f}{\sqrt{2}}c_{\beta-\alpha},\label{hff}\\
y_{Hij}^f&=\frac{\lambda_i^f}{\sqrt{2}}\delta_{ij}c_{\beta-\alpha}-\frac{\rho_{ij}^f}{\sqrt{2}}s_{\beta-\alpha}, \\
y_{Aij}^f&=\mp\frac{i\rho_{ij}^f}{\sqrt{2}},\label{yAij}
\end{align}
where $i, j$ are flavor indices, $\lambda_i^f=\sqrt{2}m_i^f/v~(v=\sqrt{v_1^2+v_2^2}=246~\text{GeV})$, 
$s_{\beta-\alpha}=\sin(\beta-\alpha)$ and $c_{\beta-\alpha}=\cos(\beta-\alpha)$ with $\alpha$ being the mixing 
angle between $h$ and $H$ while $\beta=\tan^{-1}(v_2/v_1)$.
The negative (positive) sign in Eq.~(\ref{yAij}) is for the up (down)-type fermions. 
The $3\times 3$ matrices $\rho^f$ are in general complex and can break CP explicitly
and/or induce the flavor-changing processes.
Note that the Yukawa coupling for $h$ is reduced to the SM in the limit of $c_{\beta-\alpha}\to0$ (alignment limit).
In the current study, we consider the case in which $\rtt^u$, $\rho_{bb}^d$, and 
$\rho_{ee}^e$ are non-zero and set all other $\rho_{ij}=0$ for simplicity. 
Furthermore, $\rho_{tt}^u$ is assumed to be real (for a complex $\rho_{tt}$ case, see Ref.~\cite{Fuyuto:2017ewj}).
As discussed below, the nonzero $\rho_{ee}^e$ 
plays a pivotal role in realizing a cancellation mechanism in electric dipole moment (EDM) of electron~\cite{Fuyuto:2017ewj}.
Hereafter, we omit the superscripts of $\rho$'s for notational simplicity.

As demonstrated in Refs.~\cite{Guo:2016ixx,Fuyuto:2017ewj}, with a specific ansatz for $Y_{1,2}$  $\rho_{bb}$ is given by
\footnote{Since the exemplified Yukawa ansatz leads to massless strange quark, we do not use it 
in our numerical calculation and take more realistic Yukawa ansatz.}
\begin{align}
\irbb = -\frac{1}{\lambda_b} \text{Im}[(Y_1)_{bs}(Y_2)_{bs}^*].\label{rhobb}
\end{align}
Therefore, $\rho_{bb}$ is correlated with the $b$-$s$ changing interactions in the symmetric phase,
where the Higgs VEVs are zero.  This correlation is also confirmed in a basis-invariant manner in Ref.~\cite{Guo:2016ixx}.
Since we consider the VEVs as the small perturbation in calculating the BAU (VEV insertion approximation~\cite{CTP1}),
the CP-violating source term arising from the $b$-$s$ transitions takes the form
\begin{align}
S_{\text{CPV}}  = C_{\text{BAU}}\text{Im}[(Y_1)_{bs}(Y_2)_{bs}^*],
\end{align}
where $C_{\text{BAU}}$ denotes a dynamical factor for the scattering processes
among the bottom/strange quarks and bubble wall (for the explicit form, see Refs.~\cite{Chiang:2016vgf,Fuyuto:2017ewj}).
While this baryogenesis mechanism is the same as in Ref.~\cite{Liu:2011jh}, 
the correlation of Eq.(\ref{rhobb}) is unclear in \cite{Liu:2011jh}, leading to different phenomenological 
consequences. More explicitly, the BAU-related CP violation
seems correlated with $\rho_{bs}$ rather than $\rho_{bb}$ so that there is no severe EDM constraints,
which is in stark contrast to our case  and other work~\cite{Cline:2011mm,Guo:2016ixx,Chiang:2016vgf,Fuyuto:2017ewj}. 
In principle, $\rho_{bs}$ EWBG could be possible as is the case of $\rho_{tc}$ 
EWBG discussed in Ref.~\cite{Fuyuto:2017ewj}. To this end, however, $\rho_{bs}$ has to be $\mathcal{O}(1)$ in magnitude, which is not allowed experimentally.

We calculate the BAU using closed-time-path formalism applied in 
Refs.~\cite{Guo:2016ixx,Liu:2011jh,Chiang:2016vgf,Fuyuto:2017ewj,CTP1,CTP1basedBAU1}
\footnote{While a lot of efforts have been made in developing the BAU calculation using closed-time-path
formalism~\cite{Guo:2016ixx,Liu:2011jh,Chiang:2016vgf,Fuyuto:2017ewj,deVries:2017ncy,CTP1,CTP2,CTP3,CTP1basedBAU1,CTP1basedBAU2}, 
there still exist theoretical challenges that prevent one from obtaining the robust value
(for a review, see, {\it e.g}, Ref.~\cite{Morrissey:2012db}).Theoretical uncertainties are addressed when interpreting our results.}.
The relevant particle number densities in our scenario are 
$\{Q_3 = n_{t_L}+n_{b_L}$, $T=n_{t_R}$, $B=n_{b_R}$, $S=n_{s_R}$,
$H=n_{H^+_1}+n_{H^0_1}+n_{H^+_2}+n_{H^0_2}\}$,
which are expanded to the leading order in the chemical potential $\mu$ as $n_{b,f}=T^2\mu k_{b,f}/6$,
with $b~(f)$ being bosons (fermions). One finds that $k_{b(f)}=2(1)$ in the massless limit.
The coupled diffusion equations for those number densities in the plasma frame are given by
\begin{align}
\partial_\mu j_{Q_3}^\mu 
&= -\Gamma_{Y_t}(\xi_{Q_3}+\xi_H-\xi_T)-\Gamma_{M_t}^-(\xi_{Q_3}-\xi_T) \nonumber\\
&\quad-2\Gamma_{ss}N_5+S_{b_L}, \\
\partial_\mu j_T^\mu 
&= \Gamma_{Y_t}(\xi_{Q_3}+\xi_H-\xi_T)
	+\Gamma_{M_t}^-(\xi_{Q_3}-\xi_T) \nonumber\\
&\quad+\Gamma_{ss}N_5, \\
\partial_\mu j_H^\mu 
& = -\Gamma_{Y_t}(\xi_{Q_3}+\xi_H-\xi_T)+\Gamma_{Y_{bs}}(\xi_{Q_3}-\xi_H-\xi_S) \nonumber\\
&\quad -\Gamma_H\xi_H, \\
\partial_\mu j_B^\mu&=\Gamma_{ss}N_5, \\
\partial_\mu j_S^\mu&=\Gamma_{ss}N_5-S_{b_L}, 
\end{align}
where $\xi_i=n_i/k_i$, $N_5 = 2\xi_{Q}-\xi_T-\xi_S-8\xi_B$, and
$\partial_\mu j^\mu_i=\dot{n}_i-D_i\nabla^2 n_i$ with $D_i$ denoting a diffusion constant. 
$S_{b_L}$ denotes the CP-violating source term induced by $(Y_{1,2})_{bs}$ while 
$\Gamma_{Y_t}$, $\Gamma_{M_t}^-$, $\Gamma_H$ and $\Gamma_{ss}$ are the rates by top-Higgs interactions, top-bubble wall interactions, Higgs number-violating interactions 
and strong sphaleron~\cite{StrongSph}, respectively. 
Since $\Gamma_{Y_t}, \Gamma_{\rm ss}\gg \Gamma_{M_t}^-$
the above coupled equations can be reduced to a single differential equation for $H$
~\cite{Liu:2011jh,Huet:1995sh,CTP1basedBAU2}: $\dot{H}-\bar{D}\nabla^2 H+\bar{\Gamma}H-\bar{S}+\mathcal{O}(1/\Gamma_{ss}, 1/\Gamma_{Y_t})=0$,
where $\bar{S}=k_H(k_{Q_3}-7k_T+k_B)S_{b_L}/(a+b)$ with $a=k_H(9k_{Q_3}+9k_T+k_B)$
and $b=9k_{Q_3}k_T+k_{Q_3}k_B+4k_Tk_B$ (for $\bar{D}$ and $\bar{\Gamma}$, see Ref. \cite{Liu:2011jh}). 
After transforming from the plasm frame to the wall rest frame ($z\to \bar{z}=z+v_wt$ with $v_w$ being the bubble wall velocity),
$S_{b_L}(\bar{z})\propto v_w\Delta\beta/L_w$, 
where $\Delta\beta$ is a variation of $\beta$ during the EWPT and $L_w$ the bubble wall width.

One can find the total left-handed number density as
\begin{align}
n_L(\bar{z})\simeq
\frac{r_2v_w^2}{\Gamma_{ss}\bar{D}}
\left(
	1-\frac{D_q}{\bar{D}}
\right)H(\bar{z})+\mathcal{O}(1/\Gamma_Y),\quad 
\end{align}
where $r_2=k_Hk_B^2(5k_{Q_3}+4k_T)(k_{Q_3}+2k_T)/a^2$, $D_q$ is the quark diffusion constant.
Assuming that $\bar{\Gamma}(\bar{z})$ is nonzero and constant for $\bar{z}>0$, one gets $H(\bar{z}) \simeq 
e^{v_w\bar{z}/\bar{D}}k_HL_wS_{b_L}\sqrt{a}/\sqrt{(\Gamma_{M_t}^-+\Gamma_H)\big(k_H(a+b)\bar{D}\big)}$,
where we also take the limits of $4\bar{D}\bar{\Gamma}\gg v_w^2$ and $L_w\sqrt{\bar{\Gamma}/\bar{D}}\ll 1$.
To leading order in our calculation, the $L_w$ dependence in $H(\bar{z})$ dependence drops out since $S_{b_L}\propto 1/L_w$.

After solving a diffusion equation for the baryon number density ($n_B$)~\cite{CTP2,CTP1basedBAU1,BAU_WKB}, one finds
\begin{align}
n_B = \frac{-3\Gamma_B^{(\text{sym})}}{2D_q\lambda_+}
\int_{-\infty}^0dz'~n_L(z')e^{-\lambda_-z'},\label{nB}
\end{align}
with $\lambda_\pm = \big[v_w\pm\sqrt{v_w^2+4\mathcal{R}D_q}\big]/2D_q$, 
$\Gamma_B^{(\text{sym})}$ is the $B$-changing rate via sphaleron
in the symmetric phase and $\mathcal{R}=15\Gamma_B^{(\text{sym})}/4$.

One comment on an approximation adopted in Ref.~\cite{Liu:2011jh} is that 
the CP-conserving source term induced by $(Y_{1,2})_{bs}$ is treated as the next-to-leading order
due to the fact that it is smaller than the corresponding term induced by the top quark, and thus neglected.
However, naively, the numerical impact of such a term may not be negligibly small. 
If so, the BAU based on Ref.~\cite{Liu:2011jh} would be overestimated. In our numerical analysis, 
we regard the dropped term as the part of the theoretical uncertainties and defer the improvement of the BAU calculation to future work.

Note that EWBG becomes ineffective if $v_w$ approaches to zero or gets bigger than about the speed of sound in the plasma ($1/\sqrt{3}\simeq 0.58$).
In Ref.~\cite{Ahmadvand:2013sna,Dorsch:2016nrg}, however, it is found that $0.1\lesssim v_w\lesssim 0.6$ in the softly-$Z_2$ broken 2HDMs,
where the stronger EWPT corresponds to larger $v_w$.
Since there is no serious study on $v_w$ in the G2HDM, we take $v_w=0.4$ as a reference value.
For numerical estimate of $n_B$, we take $D_q=8.9/T$ and $\Gamma_B^{(\text{sym})}=5.4\times 10^{-6}T$ 
and $\Gamma_{ss}=3.2\times 10^{-3}T$ with $T$ being temperature.

We find the BAU-viable regions by requiring that $Y_B=n_B/s$ should be greater than the
observed value $Y_B^{\text{obs}}=8.59\times 10^{-11}$~\cite{Ade:2013zuv}, where $s$ denotes the entropy density. 

The BAU can survive after the EWPT if the $B$-changing process is sufficiently suppressed. 
The rough criterion of the $B$ preservation is given $v_C/T_C\gtrsim1$, where $T_C$ denotes a critical
temperature and $v_C$ is the Higgs VEV at $T_C$. 
In our numerical analysis, we calculate $v_C/T_C$ using a finite-temperature one-loop effective potential 
with thermal resummation. 

\section{Experimental constraints}\label{expc}
Before showing the numerical results, we first outline the experimental constraints relevant to our study.
The $\rbb$ coupling is constrained by several existing measurements such as Higgs signal strengths,
branching ratio of $B\to X_s \gamma$ ($\bsg$), EDM and the asymmetry of the CP asymmetry 
between charged and neutral $\bsg$ decay ($\dcp$).

First we consider constraints from Higgs signal strength measurements.
The presence of non-zero $\cba$ and $\rho_{ij}$ modify the $h$ boson couplings
$y_{hff}$, as can be seen from Eq.\eqref{hff}. As a result $\rho_{bb}$ receives stringent constraint
if $\cba$ is non-zero. For our analysis we incorporate the Run-2 combined measurements of Higgs boson couplings by CMS~\cite{Sirunyan:2018koj}.
The result is based on  $\sqrt{s}=13$ TeV $pp$ collision with 35.9 \fbi (2016 data) and summarizes different
signal strengths $\mu_i^f$ for a specific decay mode $i\to h \to f$. The signal strength $\mu_i^f$ is defined as
\begin{align}
 \mu_i^f = \frac{\sigma_i\mathcal{B}^f}{(\sigma_i)_{\text{SM}}(\mathcal{B}^f)_{\text{SM}}} = \mu_i \mu^f,
\end{align}
where $\sigma_i$ is the production cross section for $i\to h$ and $\mathcal{B}^f$ is the branching ratio
for $h\to f$, with $i=ggF,~VBF,~Zh,~Wh,~tth$ and $f= \gamma\gamma,~ZZ,~WW,~\tau\tau,~bb,~\mu\mu$. 
We follow Refs.~\cite{Djouadi:2005gi,Branco:2011iw,Fontes:2014xva,Hou:2018uvr} for the expressions of different $\mu_i^f$.
In particular, we take two production modes, gluon fusion ($ggF$) and vector boson fusion ($VBF$)
in our analysis. We find that for the $ggF$ category, the sensitive decay modes are
$\mu_{ggF}^{\gamma\gamma}$, $\mu_{ggF}^{ZZ}$, $\mu_{ggF}^{WW}$ and $\mu_{ggF}^{\tau\tau}$, while $\mu_{VBF}^{\gamma\gamma}$,
$\mu_{VBF}^{WW}$ and $\mu_{VBF}^{\tau\tau}$ for $VBF$; these can be found from Table.~3 of Ref.~\cite{Sirunyan:2018koj}. 
Additionally, we also consider the recent observation of $h\to b\bar b$ in 
$Vh$ production by ATLAS~\cite{Aaboud:2018zhk} and CMS~\cite{Sirunyan:2018kst}. 
In order to determine the constraint on $\rbb$, we combine all these measurements and refer them together as 
``Higgs signal strength measurements''.

We now turn our attention to $\bsg$ constraint. $\bsg$ receives contribution from charged Higgs and top quark loop,
which modifies the leading order (LO) Wilson coefficient $C^{(0)}_{7,8}$ at the matching scale $\mu$. At the matching scale
$\mu = m_W$ the LO Wilson coefficients are defined as
\begin{align}
C^{(0)}_{7,8}(m_W)= F^{(1)}_{7,8}(x_t)+\delta C_{7,8}^{(0)}(\mu_W),\label{c78}
\end{align}
where $x_t=(\overline{m}_t(m_W)/m_W)^2$, $\overline{m}_t(m_W)$ $\overline{\mbox{MS}}$ running mass
of top at $m_W	$, and $F^{(1)}_{7,8}(x)$ can be found in the Ref.~\cite{Ciuchini:1997xe} (see also 
Ref.~\cite{Chetyrkin:1996vx}). The second term in Eq.\eqref{c78} arise from 
the charged Higgs contribution, which is, at LO, expressed as~\cite{Altunkaynak:2015twa}
\begin{align}
 \delta C_{7,8}^{(0)}(m_W)\simeq &\frac{|\rtt|^2}{3\lambda_t^2}F^{(1)}_{7,8}(y_{H^+}) 
 -\frac{\rtt\rho_{bb}}{\lambda_t\lambda_b}F^{(2)}_{7,8}(y_{H^+})\label{c78np},
\end{align}
with $y_{H^+}=(\overline{m}_t(m_W)/m_{H^+})^2$, while the expression for $F^{(2)}_{7,8}(y_{H^+})$ 
are given in Ref.~\cite{Ciuchini:1997xe}.
In order to find constraint on $\rho_{bb}$, we follow the prescription of Ref.~\cite{Crivellin:2013wna} and define
\begin{align}
 R_{\text{exp}}=\frac{\bsg_{\text{exp}}}{\bsg_{\text{SM}}}.
\end{align}
The current world average of $\bsg_{\text{exp}}$ extrapolated to photon energy cut $E_0=1.6$ GeV 
is $(3.32\pm0.15)\times 10^{-4}$~\cite{Amhis:2016xyh}, while the
next-to-next-to LO prediction in SM for the same photon 
energy cut is $\bsg_{\text{SM}}=(3.36\pm0.23)\times 10^{-4}$~\cite{Czakon:2015exa}.
We then demand $R_{\text{theory}}=\bsg_{\text{G2HDM}}/\bsg_{\text{SM}}$ based on our LO calculation. 
We take the matching scale and low-energy scale as $m_W$ and $\overline{m}_b(m_b)$ respectively, 
and demand $R_{\text{theory}}$ does not exceed $2\sigma$ error of $R_{\text{exp}}$.

Recently, ACME Collaboration put a new constraint on electron EDM ($d_e$), $|d_e|<1.1\times 10^{-29}~e~\text{cm}$~\cite{Andreev:2018ayy}, 
which is the most sensitive constraint on $\irbb$. 
\footnote{ We have confirmed that neutron and Mercury EDMs in our scenario are 
smaller than the current experimental bounds~\cite{Baker:2006ts,Graner:2016ses} by two- and one-order magnitude, respectively,
where the estimates are based on Refs.~\cite{Hisano:2015rna} and \cite{Cheung:2014oaa}.
Note that a cancellation scenario described below does not change this situation. }
As widely studied, the two-loop Barr-Zee diagrams~\cite{Barr:1990vd} are the leading contributions to $d_e$ in the 2HDM~\cite{EDM_2HDM}.
It is found that our $\rho_{bb}$-EWBG scenario would be virtually excluded by the new $d_e$ bound
unless the cancellation mechanism or the alignment limit are invoked~\cite{Fuyuto:2017ewj}. 
In the former case, for example, one gets $|d_e|=1.8\times 10^{-29}~e~{\text{cm}}$ for $\text{Im}\rho_{bb}=0.1$.
This can be made smaller than the current experimental upper bound
by turning on $\rho_{ee}$ as $\text{Re}\rho_{ee}=0$ and $0.06\lesssim \text{Im}\rho_{ee}/(\lambda_e\lambda_b)\lesssim 0.3$ 
that induce other Barr-Zee diagrams with the opposite sign.
In the latter case, all the EDM contributions are simply decoupled. 
In what follows, we assume the former in which phenomenological consequences are rich.

The direct CP asymmetry $\mathcal{A}_{\text{CP}}$~\cite{Kagan:1998bh} of $B\to X_s \gamma$ also
offers a very sensitive probe for $\irbb$. However, it has been proposed~\cite{Benzke:2010tq} that $\dcp$,
{\it i.e.} the asymmetry of the CP asymmetry for the charged and neutral $B\to X_s \gamma$ decay 
is even more powerful for probing CP violating effects. $\dcp$ is defined as~\cite{Benzke:2010tq}
\begin{align}
 \dcp = \mathcal{A}_{B^-\to X_s^- \gamma} - \mathcal{A}_{B^0\to X_s^0 \gamma}
 \approx 4 \pi^2 \alpha_s \frac{\tilde{\Lambda}_{78}}{m_b}\mbox{Im}\bigg(\frac{C_8}{C_7}\bigg),\label{acp}
\end{align}
where $\tilde{\Lambda}_{78}$ is a hadronic parameter, $\alpha_s$ is the strong
coupling constant at $\overline{m}_b(m_b)$. 
Recently, Belle experiment reported that $\dcp=(+3.69\pm2.65\pm0.76)\%$~\cite{Watanuki:2018xxg}~\footnote{
We are grateful to Akimasa Ishikawa for pointing out the changed central value and errors of $\dcp$ 
in the latest version of Ref.~\cite{Watanuki:2018xxg}.}, where the first uncertainty is 
statistical and the second one is systematic. In order to find the excluded region for $\rbb$, we utilize Eq.~\eqref{acp}, 
and allow $2\sigma$ error on the measured value of $\dcp$. 
In finding the constraint, we have utilized the LO Wilson coefficients as in Eq.~\eqref{c78} as first approximation.
The hadronic parameter $\tilde{\Lambda}_{78}$ is expected to be $\sim\Lambda_{\text{QCD}}$, 
and estimated to be in the range of $ 17~\mbox{MeV}<\tilde{\Lambda}_{78}<190$ MeV~\cite{Benzke:2010tq}.
In our analysis we take the average value of $\tilde{\Lambda}_{78}=89$ MeV as a reference value. 
We remark that this constraint heavily depends on the value of $\tilde{\Lambda}_{78}$ and becomes 
weaker for the smaller values of $\tilde{\Lambda}_{78}$.

\section{Results}\label{res}
\begin{figure*}[htbp!]
\center
\includegraphics[width=.385 \textwidth]{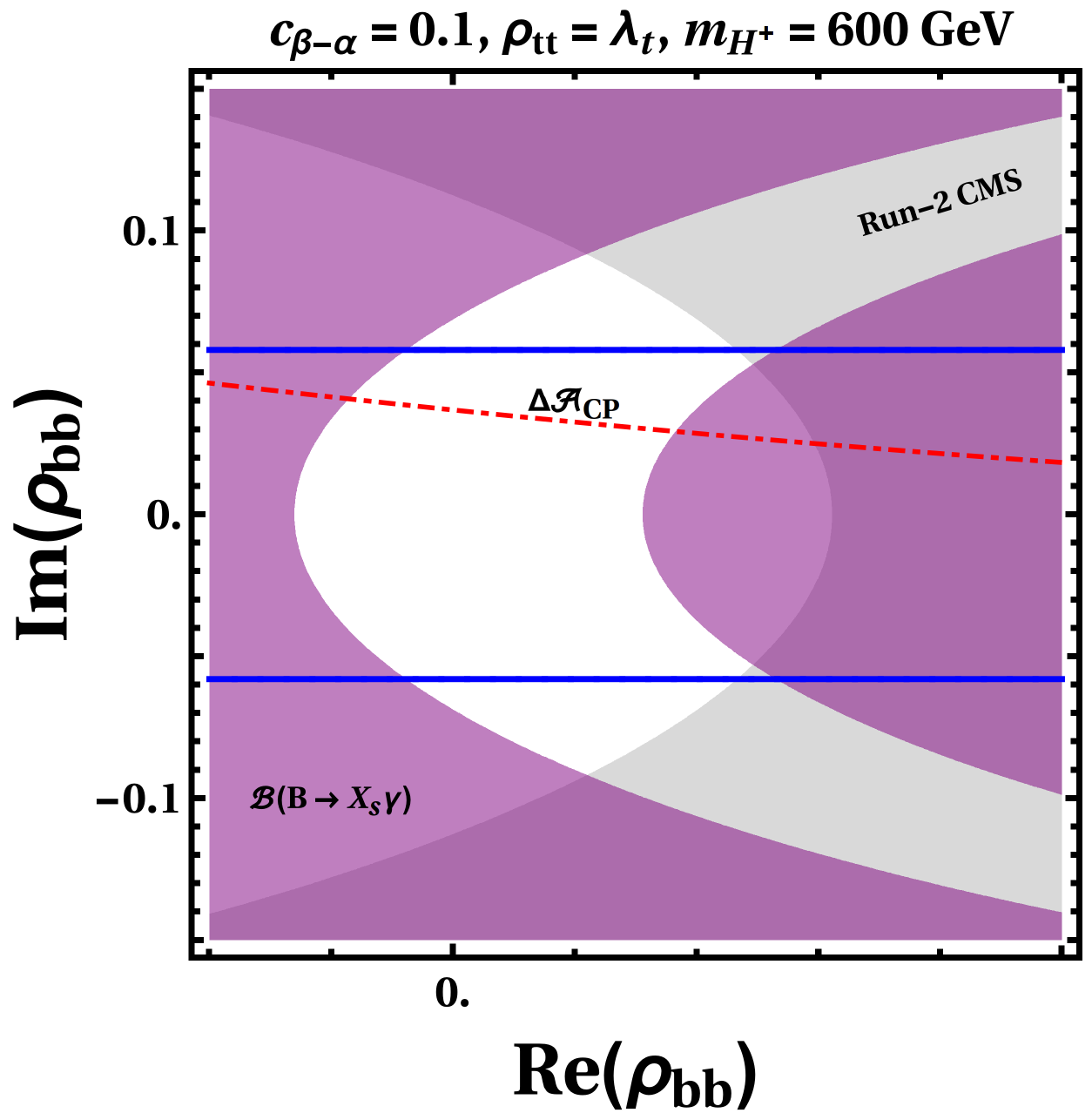}
\hspace{5mm}
\includegraphics[width=.385 \textwidth]{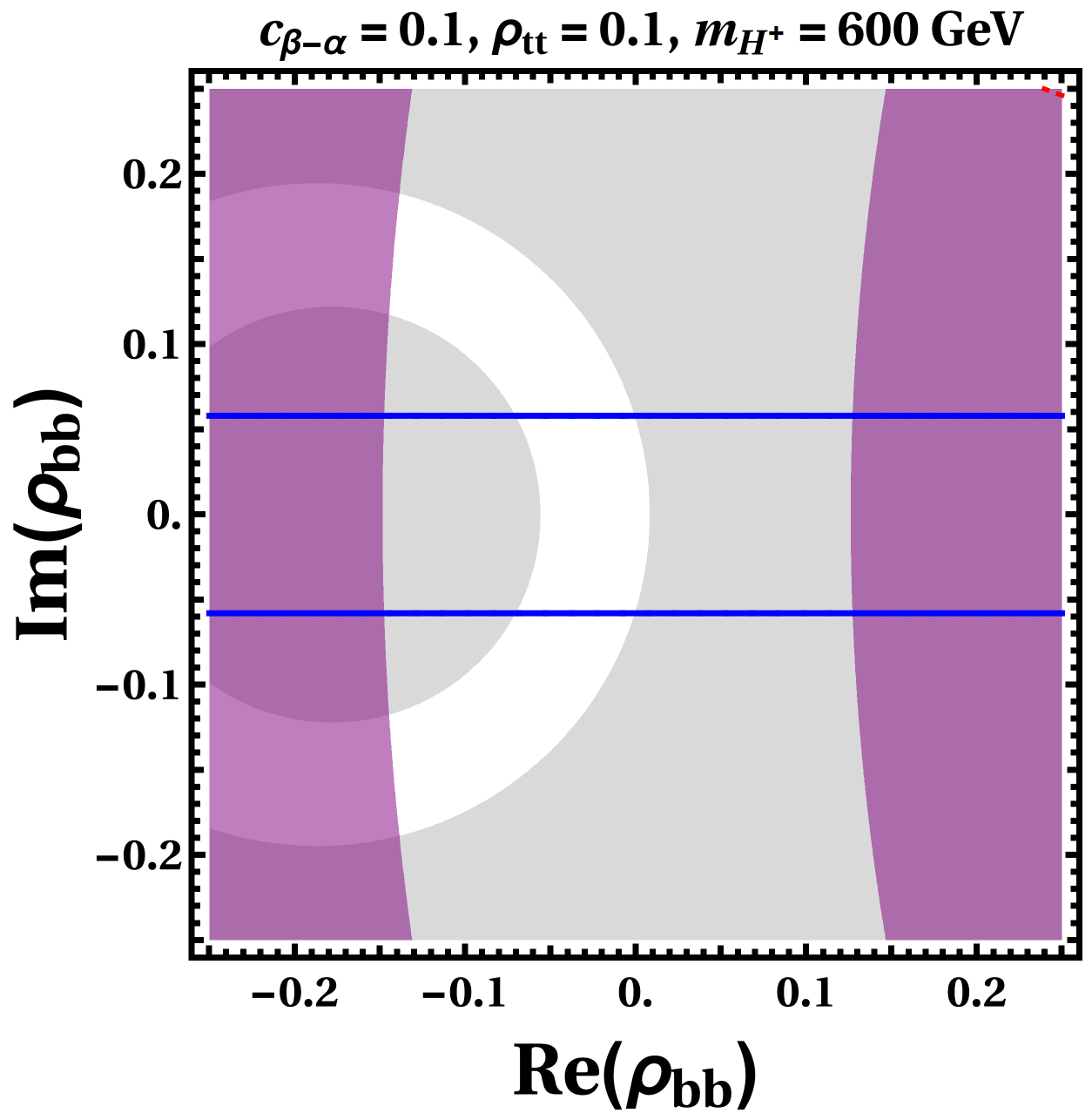}
\caption{$Y_B/Y_B^{\text{obs}}=1$ contours (blue solid contours) and the 2$\sigma$ excluded limits of the Higgs
signal strengths (gray shaded regions), $\mathcal{B}(B\to X_s\gamma)$ (purple shaded regions) and $\Delta\mathcal{A}_{\text{CP}}$ 
(red dash-dotted curves) are shown, respectively. 
We take $c_{\beta-\alpha}=0.1$, $m_H=m_A=m_{H^\pm}=600$ GeV, $\rtt=\lambda_t$ (left) and $\rtt=0.1$ (right). 
} 
\label{BAU}
\end{figure*}

For illustration we set $\cba =0.1$ and assume that $m_H=m_A=m_{H^\pm}=600$ GeV, however the impact 
of other choices will be discussed later part of this paper.
Furthermore, we take $\tan\beta=1$ and $M=400$ GeV, where $M$ is a mixing mass parameter of the
two Higgs doublet in a generic basis. This parameter choice corresponds to $\lambda_1 = 4.7$, 
$\lambda_2 = 2.4$, $\lambda_3 = 3.7$, $\lambda_4=\lambda_5= -3.3$ and $\lambda_{6}=\lambda_7=0$
with the notation of Ref.~\cite{Chiang:2016vgf}. 
\footnote{It is found that $\lambda_1(Q)>4\pi$ at $Q=2.7$ TeV and the Landau 
pole occurs at $Q=7.4$ TeV using one-loop renormalization equations. Such a low cutoff
is a generic consequence of the strong first-order EWPT in 2HDMs~\cite{Cline:2011mm,Dorsch:2016nrg,Fuyuto:2015jha}.}
With this choice, we have $T_C = 112.4$ and $v(T_C) = 191.3$ GeV.
For the input parameters for the $Y_B$ calculation, we take the parameters employed in Refs.~\cite{Chiang:2016vgf,Fuyuto:2017ewj}.
One comment we should make here is that $Y_B$ is linearly proportional to $\Delta\beta$.
Since its numerical value is unknown in the current model, 
we infer it from the results in the minimal supersymmetric standard model, {\it i.e.}, $\Delta\beta=\mathcal{O}(10^{-4}-10^{-2})$~\cite{Moreno:1998bq}.
Note that $\Delta\beta$ tends to be suppressed in the SM-like limit which is realized by the large $m_A$ limit.
In the 2HDM, however, the SM-like limit are controlled by both the heavy Higgs spectrum and $c_{\beta-\alpha}$. 
Since we do not take the exact alignment limit ($c_{\beta-\alpha}=0$), $\Delta\beta$ would not be so suppressed 
compared to the MSSM case with the same value of $m_A$. 
With this consideration, we take $|\Delta\beta|=0.015$ as a reference value.

In Fig.~\ref{BAU}, the BAU-viable regions are shown with the current experimental constraints
discussed above. We take $\rtt=\lambda_t$ (left panel) and 0.1 (right panel), respectively.
The regions of $|\irbb|\gtrsim 0.058$ give $Y_B/Y_B^{\text{obs}}>1$, 
which are indicated by the blue solid contours. Note that the regions of 
$\text{Re}(\rho_{bb})\gtrless0$ and $\irbb\gtrless0$ correspond to $\Delta\beta\gtrless0$, respectively.
The shaded regions in gray (purple) are ruled out by the Higgs signal strength measurements ($\mathcal{B}(B\to X_s\gamma)$)  at the $2\sigma$ level, 
while the 2$\sigma$ exclusion limits of $\Delta\mathcal{A}_{\text{CP}}$ are indicated by the red dash-dotted curves 
(with the regions above the dash-dotted curve is excluded).
In our analysis, we symmetrized the errors in the Higgs signal strength measurements for simplicity.
One can see that the EWBG-viable regions are rather limited by these current experimental constraints.
For $\rtt=\lambda_t$, the regions conforming $\irbb \gtrsim 0.058$ are excluded by $\dcp$ measurement (Fig.~\ref{BAU} [left]), 
however, negative $\irbb$ can still sustain $Y_B/Y_B^{\text{obs}}>1$, but $|\irbb|$ cannot be $\gtrsim 0.1$.
Note that in Fig.~\ref{BAU} [left], the $\dcp$ constraint excludes the EWBG-viable regions for $\irbb > 0$. This is because
the non-zero and positive central value of the Belle $\dcp$ measurement~\cite{Watanuki:2018xxg} and our 
choice of real and positive $\rtt = \lambda_t$ in the left panel of Fig.~\ref{BAU}. E.g. if one chooses $\rtt = - \lambda_t$, $\dcp$ constraint would exclude
EWBG-viable regions for $\irbb < 0$, however, would allow the parameter space for $\irbb > 0$.
If $\rtt=0.1$, on the other hand, $|\irbb|$ can reach around 0.2
and the EWBG-viable regions are expanded (Fig.~\ref{BAU} [right]). Note that $\Delta\mathcal{A}_{\text{CP}}$ does not give any useful bounds in this case. 
We note in passing that if we do not assume the cancellation mechanism for $d_e$, the current bound
would exclude the regions of $|\irbb|\gtrsim 0.06$, excluding the most EWBG-viable regions.
We further remark that the current constraints in Fig.~\ref{BAU}, heavily depend on $\cba$, $\rtt$ and $m_{H^\pm}$. For example, in the 
alignment limit, the constraint from Higgs signal strength measurements {\it i.e.} gray shaded region would vanish.
This is clear from  the expression of $y^f_{hij}$ (see Eq.\eqref{hff}), where the terms proportional to $\rho_{ij}$ are
modulated by $\cba$. Moreover, $\bsg$ and $\dcp$ do not depend on $\cba$,
the constraints from them will remain even for $\cba =0$. However, these two constraints vanish if
$\rtt=0$ and/or $m_{H^\pm}$ becomes too heavy. In such special case, {\it i.e.} when $\rtt=0$ and $\cba=0$, constraint on $|\irbb|$ could be 
milder.

\begin{figure*}[t]
\center
\includegraphics[width=.4 \textwidth]{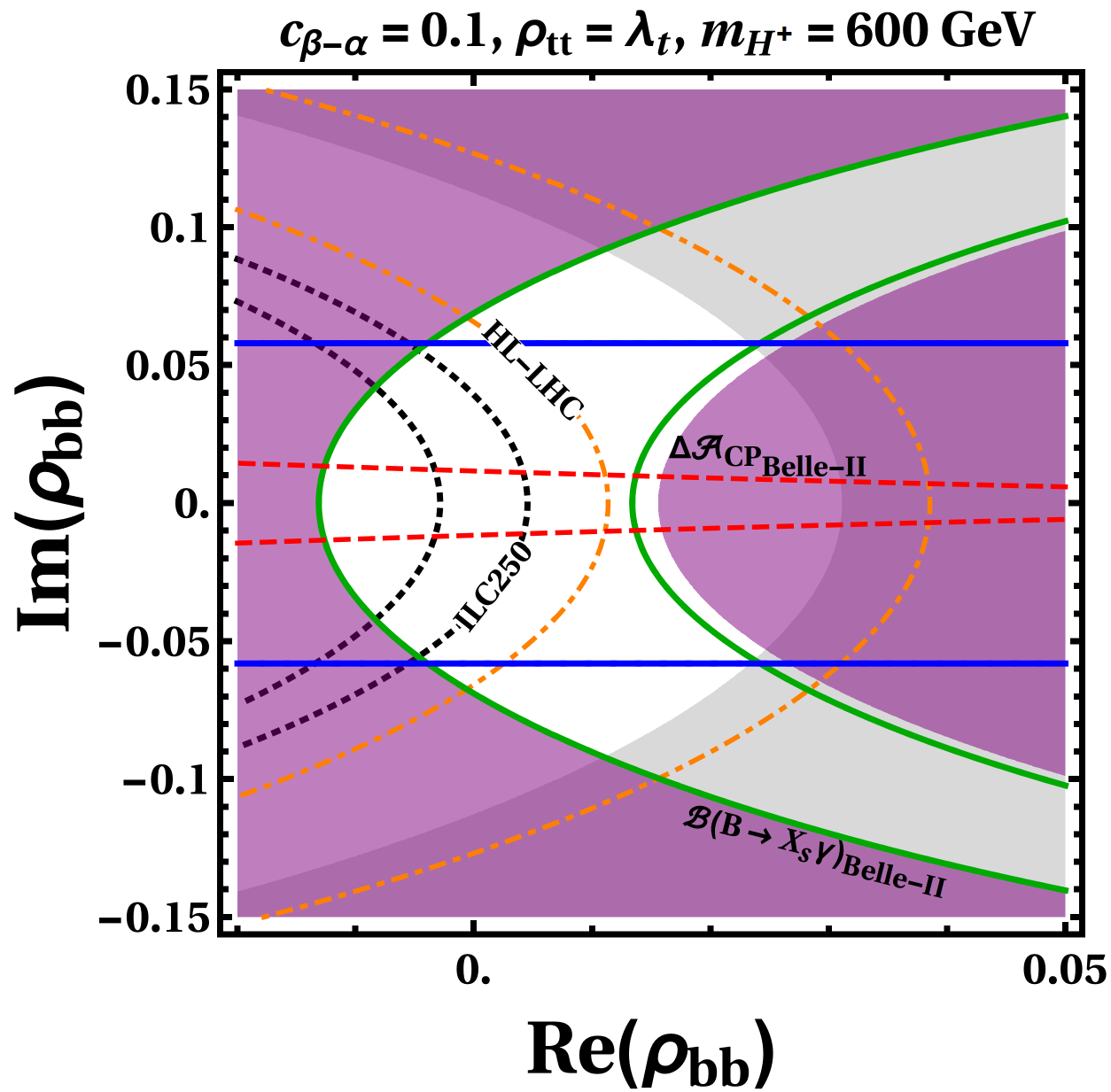}
\hspace{5mm}
\includegraphics[width=.385 \textwidth]{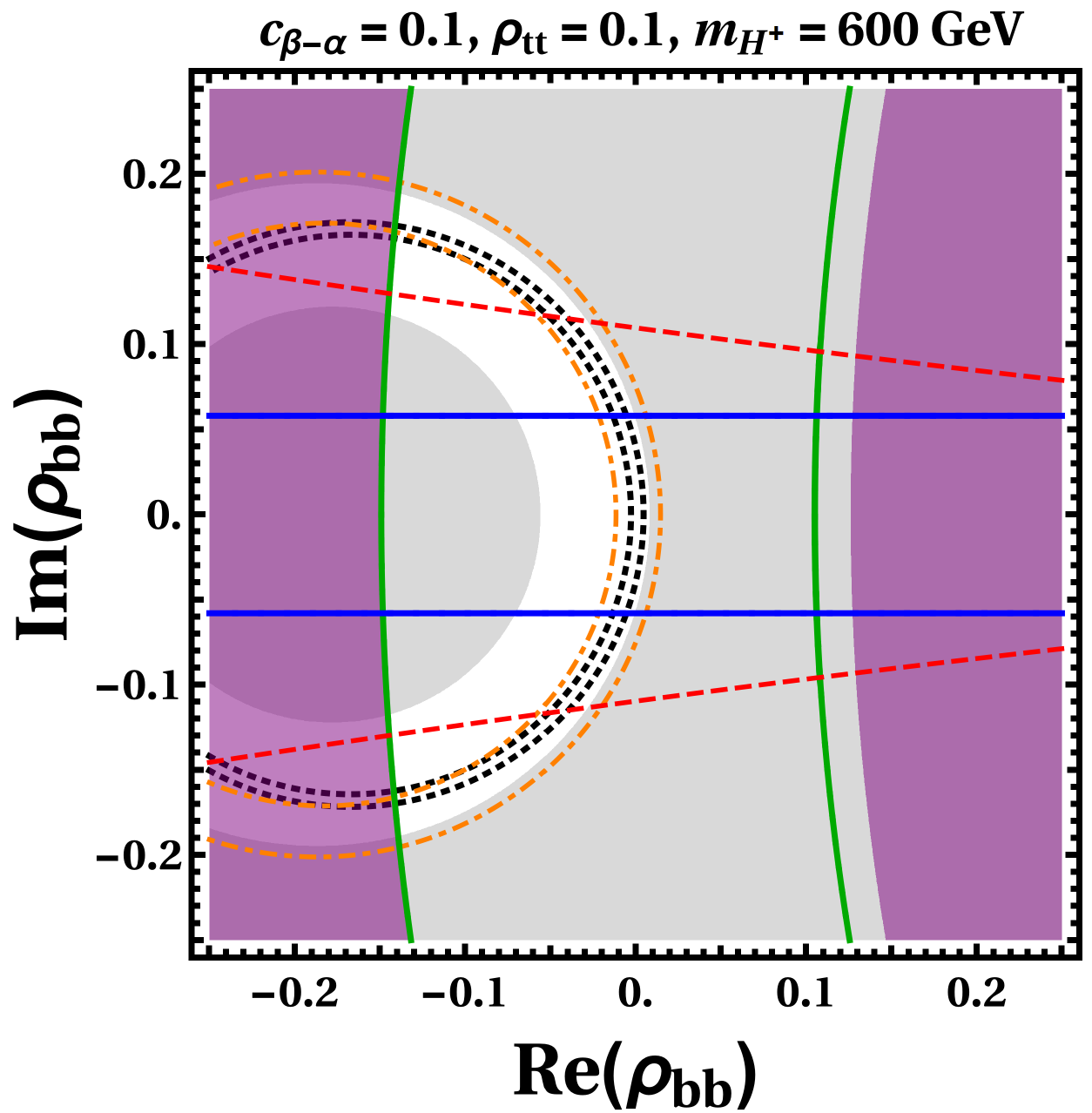}
\caption{Same as in Fig.~\ref{BAU} but future experimental sensitivities of HL-LHC (orange dash-dotted curves), 
ILC (black dotted curves) and Belle-II (green solid curve for $\mathcal{B}(B\to X_s\gamma)$ and red dotted curves
for $\Delta \mathcal{A}_{\text{CP}}$) are also overlaid. The central values for 
the future projection is assumed to be the same as in SM (Scenario-1).} 
\label{constsm}
\end{figure*}

\begin{figure*}[t]
\center
\includegraphics[width=.4 \textwidth]{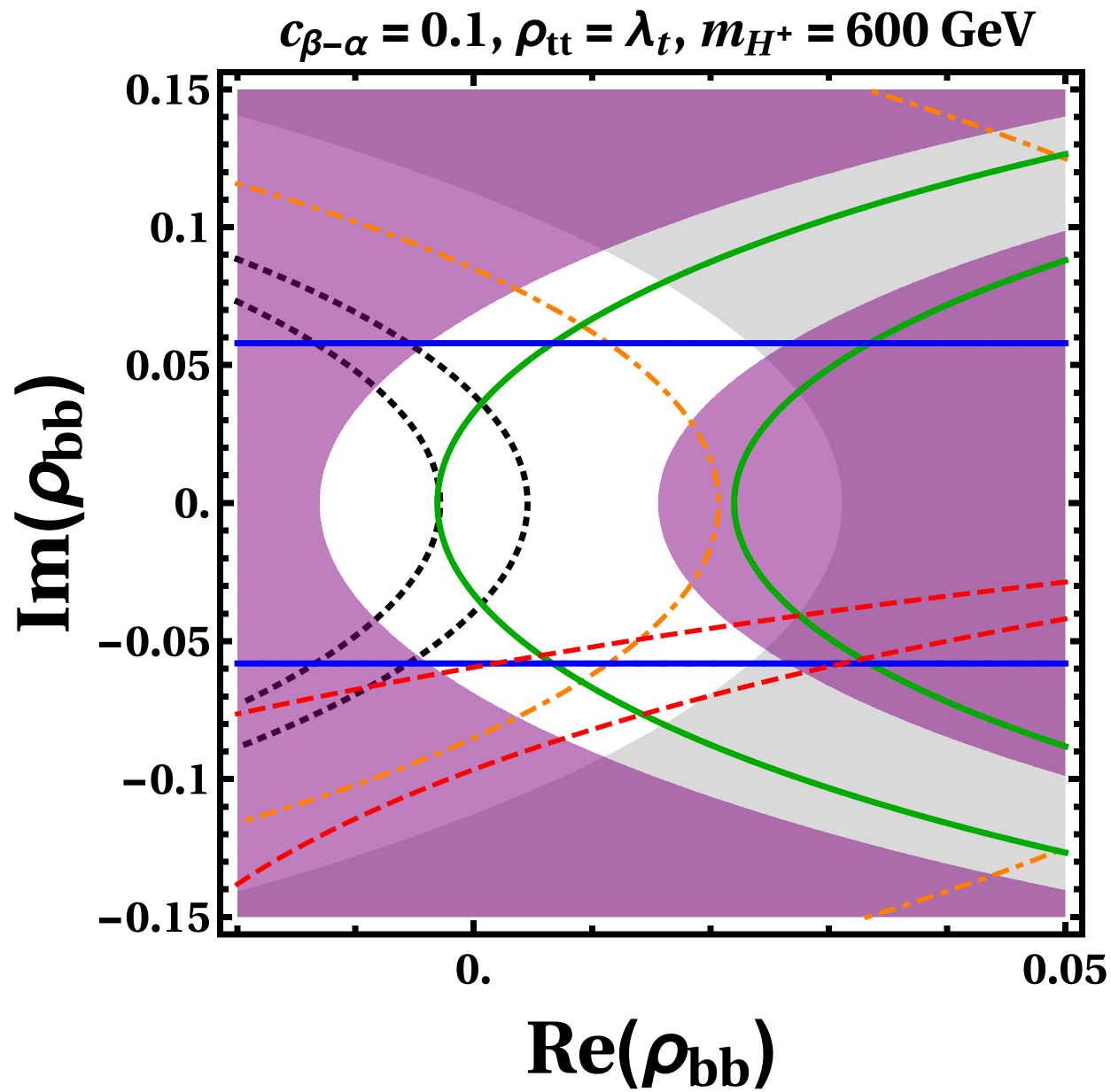}
\hspace{5mm}
\includegraphics[width=.385 \textwidth]{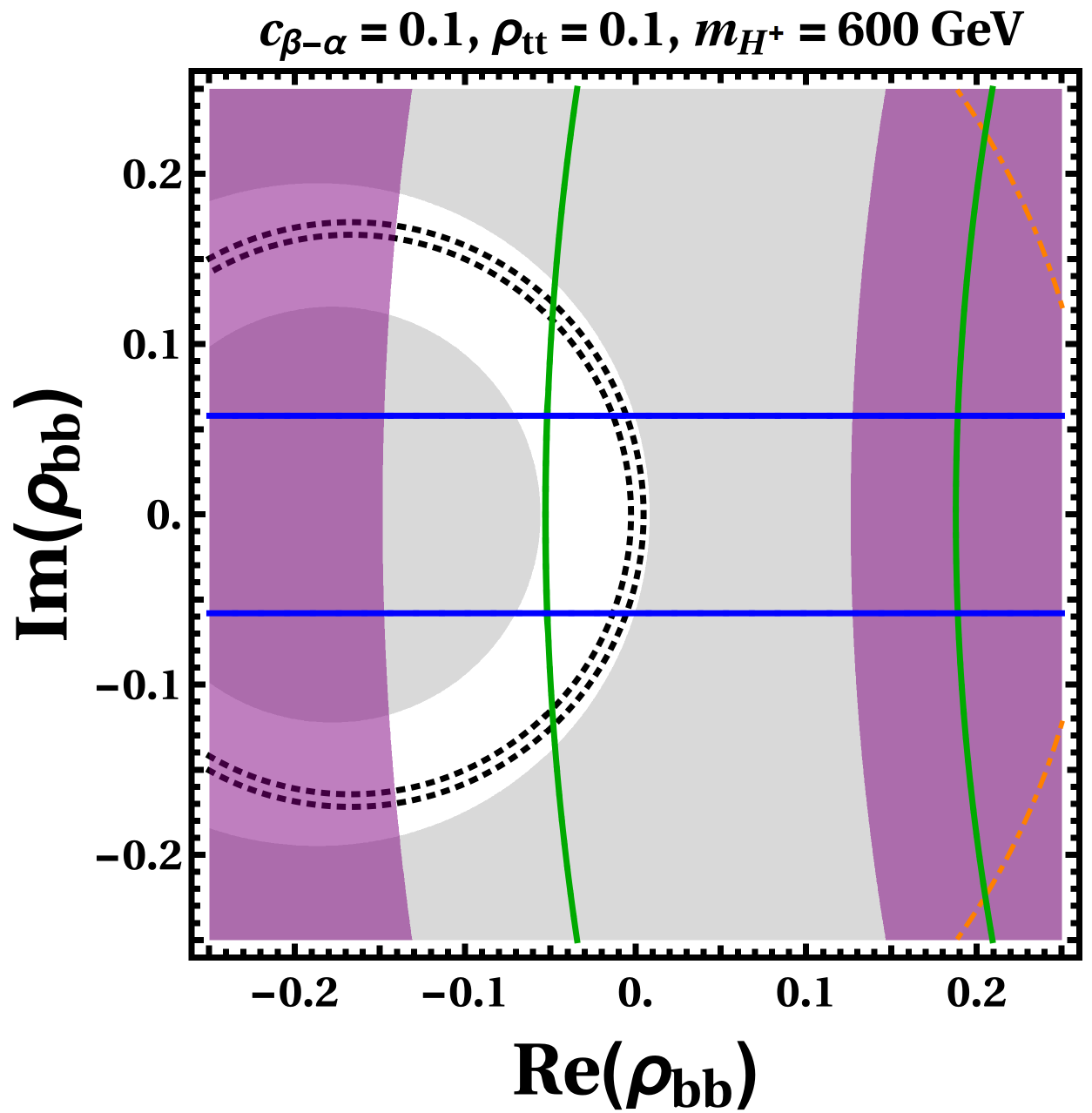}
\caption{Same as in Fig.~\ref{constsm} but the central values for the future projection
is assumed to be the same as in the current measurements (Scenario-2).} 
\label{constcrnt}
\end{figure*}

Now we discuss future prospects.
The future measurements of these observables from Belle-II, full HL-LHC dataset (3000 fb$^{-1}$) 
will also provide very sensitive probe. It will be nonetheless interesting 
to find out the parameter space for $\rbb$ assuming future projections of these constraints. In order to 
find the constraints from future projections, we adopt 
two different scenarios. In the first scenario (Scenario-1), we assume the central 
values of the future measurements for all these constraints are same
as in SM, while in the second scenario (Scenario-2) the central values are assumed to remain same as in the current measurements. 
The parameter space for $\rbb$ with the projections in Scenario-1 are summarized in Fig.~\ref{constsm}, while the projections with 
Scenario-2 are shown in Fig.~\ref{constcrnt}. 


Let us discuss the impact of these future projections in detail.
The full HL-LHC dataset is expected to measure $\mu_{ggF}^{\gamma\gamma}$, $\mu_{ggF}^{ZZ}$,
$\mu_{ggF}^{WW}$, $\mu_{VBF}^{\gamma\gamma}$ and
$\mu_{VBF}^{WW}$ very precisely, leading to very stringent constraint on $\rho_{bb}$.
For example, with an integrated luminosity of 3000 fb$^{-1}$, the projected relative uncertainties by ATLAS and
CMS~\cite{ATLAS-Higgs-projection,CMS:2017cwx}
are $\sim 5$\% for $\mu_{ggF}^{\gamma\gamma}$, $\mu_{ggF}^{ZZ}$,
$\mu_{ggF}^{WW}$, and $\sim 10$\% for  $\mu_{VBF}^{\gamma\gamma}$,
$\mu_{VBF}^{WW}$, respectively. 
We find the $2\sigma$ orange dot-dashed contours in Figs.~\ref{constsm} and \ref{constcrnt},
assuming Scenario-1 and 2, respectively.
In addition to these limits, ILC could measure~\cite{Fujii:2017vwa} the $hbb$ coupling 
at $1.1\%$ ($1\sigma$) accuracy (relative to its SM value) in the 
250 GeV program (2 ab$^{-1}$ data). We show this projected limit ($2\sigma$ exclusion) by the black dotted contours in 
Figs.~\ref{constsm} and \ref{constcrnt}.

Belle-II will also provide stringent constraint. The projected $2\sigma$ exclusion form $\bsg$ are shown
in Figs.~\ref{constsm} and \ref{constcrnt} by green solid contours, while projection for $\dcp$ is shown 
by red dashed contours. In finding these contours, we adopted similar strategy as in the HL-LHC projection of the Higgs
signal strength measurements and take two different scenarios for central values. 
For $\bsg$, we utilize the 3.2\% 
relative uncertainty for Belle-II with 50 ab$^{-1}$ data~\cite{Kou:2018nap},
in our analysis. This projected uncertainty is for the leptonic-tag $\bsg$ and is smaller than hadronic-tag or the combination of the both.
On the other hand, projected Belle-II (50 ab$^{-1}$) absolute uncertainty for $\dcp$ is 0.3\%~\cite{Kou:2018nap}.

It is clear that the future measurements offer excellent test for EWBG 
via bottom transport. These future measurements may indeed discover $\rbb$ driven EWBG. A discovery ($5\sigma$)
would be intriguing. However, it would require larger $\rbb$ compared to the projected exclusion limits ($2\sigma$)
of the different measurements as shown in Figs.~\ref{constsm} and \ref{constcrnt}.
A large part of the EWBG-viable regions can be probed by these future measurements.
For example, in Scenario-1, if $\rtt=\lambda_t$ (left panel of Fig.~\ref{constsm}), constraints from HL-LHC
(orange dot-dashed contours) and ILC-250 (black dotted contours) mutually exclude 
the regions required for $Y_B/Y_B^{\text{obs}}>1$. Additionally, in this scenario, red dashed contours from future $\dcp$ measurement
lie below $|\irbb|=0.058$. However, if  $\rtt\sim 0.1$, there exist regions where $|\irbb|\gtrsim 0.058$. 
Situation becomes completely different for Scenario-2. In this scenario, HL-LHC, ILC-250 and $\dcp$ mutually exclude
all of the regions that can support $Y_B/Y_B^{\text{obs}}>1$ both for $\rtt\sim \lambda_t$ and $\rtt\sim 0.1$.
This can be seen easily from Fig.\ref{constcrnt}.~\footnote{Note that the other orange contour lies in the right hand side
of the existing orange contour beyond the range shown in Fig.\ref{constcrnt} [right].  Besides,
the red dashed contours for the future $\dcp$ measurement lie far below $\irbb=0$.
Hence, HL-LHC, ILC-250 and future $\dcp$ mutually exclude the entire BAU-viable regions in Fig.\ref{constcrnt} [right].}
However, we stress again that the excluded regions from future projections depend on the assumptions
made on the parameters while generating Figs.~\ref{constsm} and \ref{constcrnt}. 
As discussed earlier, the constraints from HL-LHC Higgs signal strength measurements and ILC-250 vanish if $\cba =0$. 
Besides, the constraints from $\bsg$ and $\dcp$ would also vanish if
$\rtt=0$ and $m_{H^\pm}$ becomes heavy. In such scenarios, there exist finite parameter space for   
$|\irbb|$ to sustain $Y_B/Y_B^{\text{obs}}>1$.


\section{Discussions and conclusion}\label{discu} 
The interpretation of the EWBG-viable regions need some caution.
As discussed in Ref.~\cite{Chiang:2016vgf}, the BAU is subject to significant theoretical uncertainties (see also Ref.~\cite{Morrissey:2012db}). 
For example, we make use of the VEV insertion approximation that may lead to the overestimated BAU.
Likewise, as mentioned above, ignorance of the CP-conserving term induced by the $(Y_{1,2})_{bs}$
could also yield the overestimated BAU. 
In addition to those computational issues, 
impreciseness of the input parameters are also the source of the theoretical uncertainties. 
In particular, if $\Delta\beta$ is found to be one-order magnitude smaller than the value we take here, 
the BAU would get smaller by one-order magnitude, eliminating the EWBG-viable regions.  
Furthermore, there exists a discrepancy between 
the CP-violating source term calculated by our method and the one
by semi-classical force~\cite{Cline:2011mm} (see also Ref.~\cite{BAU_WKB}). 
Since the former is first order in derivative while the latter is second in derivative,
the BAU obtained by the latter scheme would become lower.  
Therefore, improvement of the BAU calculation is crucially important for the test of the scenario.
If $\rho_{bb}$ turns out to be deficient to drive the sufficient BAU in more refined calculation, 
the $\rho_{tt}/\rho_{tc}$-EWBG discussed in Ref.~\cite{Fuyuto:2017ewj} would be the
unique mechanism for baryogenesis in the G2HDM by virtue of their wider viable parameter space.
Nonetheless, the definitive conclusion cannot be made until the refined BAU calculation is available.

The constraints from $\bsg$ and $\dcp$ measurements can probe significant portion of the EWBG-viable parameter space. 
The $\dcp$ measurement with full Belle-II 50 ab$^{-1}$ dataset can probe $|\irbb| \gtrsim 0.1 $ in Scenario-1
or even can rule out entire BAU-viable region completely in Scenario-2, even for $\rtt \sim 0.1$. 
Although our assumptions on the central values for future measurements (\textit{i.e.} Scenario-1 and Scenario-2) 
are very indicative, however, we stress that the program should be revisited after the actual future measurements.
The recent measurements of $\mathcal{A}_{\text{CP}}$ and isospin violating asymmetry ($\Delta_{0+}$) 
of $B\to K^*\gamma$ decay by Belle~\cite{Horiguchi:2017ntw} may also provide complementary probe for $\irbb$,
although the theoretical predictions of these observables in general suffer from sizable uncertainties~\cite{Hurth:2010tk}.

EDM probes could come into play if their measurements are significantly improved or newly available.
For example, proton EDM, which is expected to reach $\sim10^{-29}~e~\text{cm}$ at Brookhaven
National Laboratory~\cite{Anastassopoulos:2015ura}, could give a good opportunity to confirm our 
scenario since our prediction is around $10^{-28}~e~\text{cm}$. Follow-up studies along this line are worth pursuing.

The future updates from  HFLAV for the global average of $\bsg$ would also play a major
role in constraining BAU-viable region, if $\rtt$ is not vanishingly small.  
In this regard, we remark that the $B_q-\bar{B}_q$ ($q=d,s$) mixing~\cite{Altunkaynak:2015twa} and the recent discovery of 
$t\bar t h$~\cite{Sirunyan:2018hoz,Aaboud:2018urx} would provide independent probes~\cite{Hou:2018uvr} for $\rtt$.

The Higgs signal strength measurements at HL-LHC would be complementary in probing $\rbb$
regardless of the value of $\rtt$, however $\cba$ should not be very small. 
It should be noted that $|\irbb|$ can not be too large for non-zero $\cba$. 
The current limit on the $h$ boson total width $\Gamma_h < 0.013$ GeV 
(95\% CL)~\cite{Tanabashi:2018oca} sets upper limit on $|\rbb|$ if $\cba \neq 0$.
Utilizing this limit we find that for $\rrbb=0$ and $\cba=0.1$, $|\irbb|\lesssim 0.36$ at 95\% CL. 
In determining the upper limit on $|\irbb|$ we used
LO decay width of $h$ for simplicity. We also remark, 
like Run 1 combination~\cite{Khachatryan:2016vau}, a Run 2 combined fit of ATLAS  and CMS
Higgs signal strengths would be more indicative. 
Further, our study illustrates, ILC 250 GeV run might probe $\rbb$
better than HL-LHC. It is not surprising that ILC, 
even its 250 GeV program, presents better probe 
for NP in bottom Yukawa than HL-LHC.

Also, LHC might offer direct detection of $\rbb$ driven EWBG. A non-zero $\irbb$ induces
$gg\to b \bar b A (H) \to b\bar b Z H (A)$ process if $m_A > m_H +m_Z$ ($m_H > m_A +m_Z$). 
This process provides unique probe for the EWBG, \textit{even} for $\cba=0$ and/or $\rtt =0$. 
Notwithstanding, if $\cba$ is not too small direct detection program can cover $gg\to b \bar b A\to  b \bar b Z h$. 
A discovery would be intriguing. Furthermore, for moderate values of $\rtt$, 
$ g g \to t \bar t A/t \bar tH \to t \bar t b \bar b$ with leptonic decays of at least one top
and $A/H\to b \bar b$ could be interesting. These would be studied elsewhere.

In conclusion, motivated by recent discovery of Higgs boson decay to bottom quarks,
we have analyzed the possibility of EWBG by extra bottom Yukawa $\rbb$ in the G2HDM. 
After satisfying all existing constraints, 
we found that indeed $\rbb$ can generate successful BAU, 
however, $|\irbb|$ required to be $\gtrsim 0.058$. 
For a wide range of parameter space, 
future measurements from Belle-II, Higgs signal 
strengths at HL-LHC and ILC will provide exquisite probes for such scenario. 
If the additional scalar and pseudoscalar are in the sub-TeV range,
the program can also be covered by direct searches at LHC.

\vskip0.2cm
\begin{acknowledgments}
\noindent{\bf Acknowledgments} \
We thank Wei-Shu Hou and Masaya Kohda for discussions. We also thank 
Akimasa Ishikawa for communication and discussions.
T.M. is supported by grant No. MOST-107-2811-M-002-3069 of R.O.C Taiwan and
E.S. is supported by IBS under the project code, IBS-R018-D1.
\end{acknowledgments}



\begin{thebibliography}{99}

\bibitem{Tanabashi:2018oca} 
  M.~Tanabashi {\it et al.} [Particle Data Group],
  Phys.\ Rev.\ D {\bf 98}, no. 3, 030001 (2018).

\bibitem{125h}  
  G.~Aad {\it et al.} [ATLAS Collaboration],
  Phys.\ Lett.\ B {\bf 716}, 1 (2012);~
%
  S.~Chatrchyan {\it et al.} [CMS Collaboration],
  Phys.\ Lett.\ B {\bf 716}, 30 (2012).

\bibitem{ewbg}
  V.A.~Kuzmin, V.A.~Rubakov and M.E.~Shaposhnikov,
  Phys.\ Lett.\ B {\bf 155}, 36 (1985).
For some reviews, 
see e.g.
%
  M.~Quiros,
  Helv.\ Phys.\ Acta {\bf 67}, 451 (1994);
%
  V.A.~Rubakov and M.E.~Shaposhnikov,
  Usp.\ Fiz.\ Nauk {\bf 166}, 493 (1996)
  [Phys.\ Usp.\  {\bf 39}, 461 (1996)];
%
  K.~Funakubo,
  Prog.\ Theor.\ Phys.\  {\bf 96}, 475 (1996);~
%
  A.~Riotto,
  hep-ph/9807454;~
%
%
 W.~Bernreuther,
 Lect.\ Notes Phys.\  {\bf 591}, 237 (2002);~
%
  J.M.~Cline,
  arXiv:hep-ph/0609145;~
%
  T.~Konstandin,
  Phys.\ Usp.\  {\bf 56}, 747 (2013).

\bibitem{Morrissey:2012db}
  D.E.~Morrissey and M.J.~Ramsey-Musolf,
  New J.\ Phys.\  {\bf 14}, 125003 (2012).
%
\bibitem{Carena:2004ha} 
  M.~Carena, A.~Megevand, M.~Quiros and C.~E.~M.~Wagner,
  Nucl.\ Phys.\ B {\bf 716}, 319 (2005)
%
\bibitem{Huang:2016cjm} 
  P.~Huang, A.~J.~Long and L.~T.~Wang,
  Phys.\ Rev.\ D {\bf 94}, no. 7, 075008 (2016).


\bibitem{Aaboud:2018zhk} 
  M.~Aaboud {\it et al.} [ATLAS Collaboration],
  Phys.\ Lett.\ B {\bf 786}, 59 (2018).

\bibitem{Sirunyan:2018kst} 
  A.~M.~Sirunyan {\it et al.} [CMS Collaboration],
  Phys.\ Rev.\ Lett.\  {\bf 121}, no. 12, 121801 (2018).

\bibitem{CKM}
  N.~Cabibbo,
  Phys.\ Rev.\ Lett.\  {\bf 10}, 531 (1963);~
%
  M.~Kobayashi and T.~Maskawa,
  Prog.\ Theor.\ Phys.\  {\bf 49}, 652 (1973).
  
  
\bibitem{Branco:2011iw} 
  G.~C.~Branco, P.~M.~Ferreira, L.~Lavoura, M.~N.~Rebelo, M.~Sher and J.~P.~Silva,
  Phys.\ Rept.\  {\bf 516}, 1 (2012).

\bibitem{Tulin:2011wi} 
  S.~Tulin and P.~Winslow,
  Phys.\ Rev.\ D {\bf 84}, 034013 (2011).
%
\bibitem{Cline:2011mm} 
  J.~M.~Cline, K.~Kainulainen and M.~Trott,
  JHEP {\bf 1111}, 089 (2011).

\bibitem{Guo:2016ixx} 
  H.~K.~Guo, Y.~Y.~Li, T.~Liu, M.~Ramsey-Musolf and J.~Shu,
  Phys.\ Rev.\ D {\bf 96}, no. 11, 115034 (2017).

\bibitem{Liu:2011jh} 
  T.~Liu, M.~J.~Ramsey-Musolf and J.~Shu,
  Phys.\ Rev.\ Lett.\  {\bf 108}, 221301 (2012).

\bibitem{Chiang:2016vgf} 
  C.~W.~Chiang, K.~Fuyuto and E.~Senaha,
  Phys.\ Lett.\ B {\bf 762}, 315 (2016);~

\bibitem{Fuyuto:2017ewj} 
  K.~Fuyuto, W.~S.~Hou and E.~Senaha,
  Phys.\ Lett.\ B {\bf 776}, 402 (2018).
%
\bibitem{deVries:2017ncy} 
  J.~de Vries, M.~Postma, J.~van de Vis and G.~White,
  JHEP {\bf 1801}, 089 (2018).
  
\bibitem{CTP1}
  A.~Riotto,
  Nucl.\ Phys.\ B {\bf 518}, 339 (1998);~
  A.~Riotto,
  Phys.\ Rev.\ D {\bf 58}, 095009 (1998).

\bibitem{CTP1basedBAU1}
  C.~Lee, V.~Cirigliano and M.~J.~Ramsey-Musolf,
  Phys.\ Rev.\ D {\bf 71}, 075010 (2005);~
%
  V.~Cirigliano, M.~J.~Ramsey-Musolf, S.~Tulin and C.~Lee,
  Phys.\ Rev.\ D {\bf 73} (2006) 115009;~
  %
  D.~J.~H.~Chung, B.~Garbrecht, M.~J.~Ramsey-Musolf and S.~Tulin,
  JHEP {\bf 0912} (2009) 067;~
%
  D.~J.~H.~Chung, B.~Garbrecht, M.~J.~Ramsey-Musolf and S.~Tulin,
  Phys.\ Rev.\ D {\bf 81} (2010) 063506.
  
\bibitem{CTP2}
  M.~Carena, J.~M.~Moreno, M.~Quiros, M.~Seco and C.~E.~M.~Wagner,
  Nucl.\ Phys.\ B {\bf 599} (2001) 158;~
%
  M.~Carena, M.~Quiros, M.~Seco and C.~E.~M.~Wagner,
  Nucl.\ Phys.\ B {\bf 650} (2003) 24.

\bibitem{CTP3}
  T.~Prokopec, M.~G.~Schmidt and S.~Weinstock,
  Annals Phys.\  {\bf 314} (2004) 208;~
%
  T.~Prokopec, M.~G.~Schmidt and S.~Weinstock,
  Annals Phys.\  {\bf 314} (2004) 267;~
%
  T.~Konstandin, T.~Prokopec and M.~G.~Schmidt,
  Nucl.\ Phys.\ B {\bf 716} (2005) 373;~
%
  T.~Konstandin, T.~Prokopec, M.~G.~Schmidt and M.~Seco,
  Nucl.\ Phys.\ B {\bf 738} (2006) 1.

  \bibitem{CTP1basedBAU2}
  V.~Cirigliano, C.~Lee, M.~J.~Ramsey-Musolf and S.~Tulin,
  Phys.\ Rev.\ D {\bf 81} (2010) 103503;~
 %
  V.~Cirigliano, C.~Lee and S.~Tulin,
  Phys.\ Rev.\ D {\bf 84} (2011) 056006.
  
\bibitem{StrongSph}  
  R.~N.~Mohapatra and X.~m.~Zhang,
  Phys.\ Rev.\ D {\bf 45} (1992) 2699;~
 %
  G.~F.~Giudice and M.~E.~Shaposhnikov,
  Phys.\ Lett.\ B {\bf 326} (1994) 118.
 
\bibitem{Huet:1995sh}
  P.~Huet and A.~E.~Nelson,
  Phys.\ Rev.\ D {\bf 53} (1996) 4578.

\bibitem{BAU_WKB}
  J.~M.~Cline, M.~Joyce and K.~Kainulainen,
  JHEP {\bf 0007} (2000) 018;~
%
  L.~Fromme, S.~J.~Huber and M.~Seniuch,
  JHEP {\bf 0611} (2006) 038;~
%
  L.~Fromme and S.~J.~Huber,
  JHEP {\bf 0703} (2007) 049.


  


    
\bibitem{Ahmadvand:2013sna}
  M.~Ahmadvand,
  Int.\ J.\ Mod.\ Phys.\ A {\bf 29} (2014) no.20,  1450090.
%
\bibitem{Dorsch:2016nrg}
  G.~C.~Dorsch, S.~J.~Huber, T.~Konstandin and J.~M.~No,
  JCAP {\bf 1705} (2017) no.05,  052.
  
      
\bibitem{Ade:2013zuv} 
  P.~A.~R.~Ade {\it et al.} [Planck Collaboration],
  Astron.\ Astrophys.\  {\bf 571}, A16 (2014).
    
\bibitem{Sirunyan:2018koj} 
  A.~M.~Sirunyan {\it et al.} [CMS Collaboration],
  arXiv:1809.10733 [hep-ex].
  
\bibitem{Djouadi:2005gi} 
  A.~Djouadi,
  Phys.\ Rept.\  {\bf 457}, 1 (2008).
  
\bibitem{Fontes:2014xva} 
  D.~Fontes, J.~C.~Romão and J.~P.~Silva,
  JHEP {\bf 1412}, 043 (2014).
    
\bibitem{Hou:2018uvr} 
  W.~S.~Hou, M.~Kohda and T.~Modak,
  Phys.\ Rev.\ D {\bf 98}, 075007 (2018).
  
%
  
\bibitem{Ciuchini:1997xe} 
  M.~Ciuchini, G.~Degrassi, P.~Gambino and G.~F.~Giudice,
  Nucl.\ Phys.\ B {\bf 527}, 21 (1998).

\bibitem{Chetyrkin:1996vx} 
  K.~G.~Chetyrkin, M.~Misiak and M.~Munz,
  Phys.\ Lett.\ B {\bf 400}, 206 (1997)
  Erratum: [Phys.\ Lett.\ B {\bf 425}, 414 (1998)].
    
  
\bibitem{Altunkaynak:2015twa} 
  B.~Altunkaynak, W.~S.~Hou, C.~Kao, M.~Kohda and B.~McCoy,
  Phys.\ Lett.\ B {\bf 751}, 135 (2015)
  

  
\bibitem{Crivellin:2013wna} 
  A.~Crivellin, A.~Kokulu and C.~Greub,
  Phys.\ Rev.\ D {\bf 87}, no. 9, 094031 (2013).
  
  
\bibitem{Amhis:2016xyh} 
  Y.~Amhis {\it et al.} [HFLAV Collaboration],
  Eur.\ Phys.\ J.\ C {\bf 77}, no. 12, 895 (2017).
  
  
\bibitem{Czakon:2015exa} 
  M.~Czakon, P.~Fiedler, T.~Huber, M.~Misiak, T.~Schutzmeier and M.~Steinhauser,
  JHEP {\bf 1504}, 168 (2015).
  
\bibitem{Andreev:2018ayy} 
  V.~Andreev {\it et al.} [ACME Collaboration],
  Nature {\bf 562}, no. 7727, 355 (2018).
  
\bibitem{Baker:2006ts} 
  C.~A.~Baker {\it et al.},
  Phys.\ Rev.\ Lett.\  {\bf 97}, 131801 (2006).
%
\bibitem{Graner:2016ses} 
  B.~Graner, Y.~Chen, E.~G.~Lindahl and B.~R.~Heckel,
  Phys.\ Rev.\ Lett.\  {\bf 116}, no. 16, 161601 (2016)
  Erratum: [Phys.\ Rev.\ Lett.\  {\bf 119}, no. 11, 119901 (2017)].
 
\bibitem{Hisano:2015rna} 
  J.~Hisano, D.~Kobayashi, W.~Kuramoto and T.~Kuwahara,
  JHEP {\bf 1511}, 085 (2015).
    
\bibitem{Cheung:2014oaa} 
  K.~Cheung, J.~S.~Lee, E.~Senaha and P.~Y.~Tseng,
  JHEP {\bf 1406}, 149 (2014).
      
\bibitem{Barr:1990vd} 
  S.~M.~Barr and A.~Zee,
  Phys.\ Rev.\ Lett.\  {\bf 65}, 21 (1990)
  Erratum: [Phys.\ Rev.\ Lett.\  {\bf 65}, 2920 (1990)].
  
\bibitem{EDM_2HDM}  
  M.~Jung and A.~Pich,
  JHEP {\bf 1404}, 076 (2014);~
%
  S.~Inoue, M.~J.~Ramsey-Musolf and Y.~Zhang,
  Phys.\ Rev.\ D {\bf 89}, no. 11, 115023 (2014);~
%
  K.~Cheung, J.~S.~Lee, E.~Senaha and P.~Y.~Tseng,
  JHEP {\bf 1406}, 149 (2014);~
%
  T.~Abe, J.~Hisano, T.~Kitahara and K.~Tobioka,
  JHEP {\bf 1401}, 106 (2014)
  Erratum: [JHEP {\bf 1604}, 161 (2016)].
  
\bibitem{Kagan:1998bh} 
  A.~L.~Kagan and M.~Neubert,
  Phys.\ Rev.\ D {\bf 58}, 094012 (1998).
  
\bibitem{Benzke:2010tq} 
  M.~Benzke, S.~J.~Lee, M.~Neubert and G.~Paz,
  Phys.\ Rev.\ Lett.\  {\bf 106}, 141801 (2011).
    
\bibitem{Watanuki:2018xxg} 
  S.~Watanuki {\it et al.},
  arXiv:1807.04236 [hep-ex].
  
\bibitem{Fuyuto:2015jha}
  K.~Fuyuto and E.~Senaha,
  Phys.\ Lett.\ B {\bf 747} (2015) 152.

\bibitem{Moreno:1998bq}
  J.~M.~Moreno, M.~Quiros and M.~Seco,
  Nucl.\ Phys.\ B {\bf 526} (1998) 489.


%
\bibitem{ATLAS-Higgs-projection}
ATLAS Collaboration,
  ATL-PHYS-PUB-2014-016.
%
\bibitem{CMS:2017cwx}
  CMS Collaboration,
  CMS-PAS-FTR-16-002.
  
\bibitem{Fujii:2017vwa} 
  K.~Fujii {\it et al.},
  arXiv:1710.07621 [hep-ex].

\bibitem{Kou:2018nap}
  E.~Kou {\it et al.} [Belle II Collaboration],
  arXiv:1808.10567 [hep-ex].
%
  
\bibitem{Horiguchi:2017ntw} 
  T.~Horiguchi {\it et al.} [Belle Collaboration],
  Phys.\ Rev.\ Lett.\  {\bf 119}, no. 19, 191802 (2017).
%
\bibitem{Hurth:2010tk} 
  T.~Hurth and M.~Nakao,
  Ann.\ Rev.\ Nucl.\ Part.\ Sci.\  {\bf 60}, 645 (2010).
    
\bibitem{Anastassopoulos:2015ura} 
  V.~Anastassopoulos {\it et al.},
  Rev.\ Sci.\ Instrum.\  {\bf 87}, no. 11, 115116 (2016).
    
%
\bibitem{Aaboud:2018urx}
  M.~Aaboud {\it et al.} [ATLAS Collaboration],
  Phys.\ Lett.\ B {\bf 784}, 173 (2018).
  
%
\bibitem{Sirunyan:2018hoz}
  A.M.~Sirunyan {\it et al.} [CMS Collaboration],
  Phys.\ Rev.\ Lett.\  {\bf 120}, 231801 (2018).
%
%
\bibitem{Khachatryan:2016vau}
  G.~Aad {\it et al.} [ATLAS and CMS Collaborations],
  JHEP {\bf 1608}, 045 (2016).


\end{thebibliography}
\end{document}